\documentclass[twocolumn, 10pt, amsmath, amssymb, prl, showpacs]{revtex4}

\usepackage{epsfig, amssymb, stmaryrd, times}

\newcommand{\C}{{\mathbb{C}}}

\def\be{\begin{eqnarray}}
\def\ee{\end{eqnarray}}
\def\bee{\begin{eqnarray*}}
\def\eee{\end{eqnarray*}}

\begin{document}

\title{Classical spin models and the quantum stabilizer formalism}

\author{M. Van den Nest$^1$, W. D\"ur$^{1,2}$, and H. J. Briegel$^{1,2}$ }

\affiliation{$^1$ Institut f\"ur Quantenoptik und Quanteninformation der \"Osterreichischen Akademie der Wissenschaften, Innsbruck, Austria\\
$^2$ Institut f{\"u}r Theoretische Physik, Universit{\"a}t
Innsbruck, Technikerstra{\ss}e 25, A-6020 Innsbruck,
Austria}
\date{\today}

\date{\today}
\def\makeheadbox{}

\begin{abstract}
We relate a large class of  classical spin models,
including the inhomogeneous Ising, Potts, and clock models
of $q$-state spins on arbitrary graphs, to problems in
quantum physics. More precisely, we show how to express
partition functions as inner products between certain
quantum stabilizer states and product states. This
connection allows us to use powerful techniques developed
in quantum information theory, such as the stabilizer
formalism and classical simulation techniques, to gain
general insights into these models in a unified way. We
recover and generalize several symmetries and high-low
temperature dualities, and we provide an efficient
classical evaluation of partition functions for all
interaction graphs with a bounded tree-width.
\end{abstract}

%75.10.Hk    Classical spin models
%75.10.Pq    Spin chain models
%02.70.-c    Computational techniques
%03.67.Lx    Quantum computation

\maketitle

%---------------------------------------------------
%---------------------------------------------------
{\bf 1.--- Introduction.}
%---------------------------------------------------
%---------------------------------------------------
Classical spin models, such as the Ising, Potts and clock
models \cite{Wu83}, are widely studied in statistical
physics. Despite of their simplicity, these models show a
number of highly non-trivial features, which is  e.g.
apparent in their rich phase structure and critical
behavior. These models are considered in the context of
(anti)ferromagnetism and are related to a number of central
problems in lattice statistics. A quantity of fundamental
interest is the partition function, from which other
physically relevant quantities can be derived. General
symmetry and duality relations, as well as the efficient evaluation
of the partition function are important problems in this
context. In particular, it is known that for general
interaction patterns the evaluation of the partition
function of e.g. the Potts model is NP-hard
\cite{Ba82}, rendering the problem highly non-trivial.

In this letter, we relate general  classical spin models of
$q$-state spins that interact pairwise according to some
arbitrary interaction pattern, to problems in {\em quantum
theory}, thereby obtaining a novel approach to tackle these
problems. More precisely, we establish a correspondence
between, one the one hand, classical partition functions,
and, on the other hand, overlaps between certain quantum
states---known as \emph{stabilizer states}
\cite{Go97, He06}---and complete product states of $N$
$q$--dimensional quantum systems; here $N$ is the number of
pairwise interactions. The stabilizer states encode the
interaction pattern of the model, while the product states
encode the details of the interaction. This correspondence
allows us to use a whole body of powerful results obtained
in the context of quantum information theory regarding the
classical description of such quantum states, including the
stabilizer formalism \cite{Go97, He06, Ho05} and the
classical simulation of quantum systems \cite{Shi05, Va06}.
Conversely, the vast knowledge on classical spin models
yields insights in the possible use of the corresponding
states in quantum information tasks.

We emphasize that the current approach allows  one to
investigate classical spin models by {\em virtually} going
to quantum systems---this is not be to confused with the
solution of classical problems by implementing quantum
algorithms on a quantum computer. Despite of the fact that
the formulation of the problem in terms of quantum theory
appears to increase the complexity and the number of
involved parameters, this approach, as we will show, leads
to a fruitful way to gain insights in spin models. For
example, recently established classical algorithms
\cite{Shi05, Va06} that allow one to describe and locally
manipulate certain quantum systems in an efficient way, can
be utilized to efficiently evaluate the partition functions
of these models for all interaction patterns corresponding
to graphs with a (logarithmically) bounded tree-width
\cite{twd}. Furthermore, our approach leads to a simple way
of recovering and extending general symmetry and duality
relations of the above models in a unified way. For
instance, using the fact that the corresponding states are
stabilizer states, one can relate the stabilizer groups of
the states to symmetry groups of the corresponding
partition functions. In the following we will first
describe the general class of classical spin models we
consider, and then formulate the problem in a quantum
language and apply techniques established in the context of
quantum mechanics to gain insight in the classical models.

\

%---------------------------------------------------
%---------------------------------------------------
{\bf 2.--- Classical spin models.}
%---------------------------------------------------
%---------------------------------------------------
We will consider a general class of classical
spin models on arbitrary graphs. We consider $n$ classical spins $\{s_i\}$ that can assume $q$
possible states, $s_i\in\{0, \dots, q-1\}$, and which interact pairwise according to an interaction pattern specified by a graph. More precisely, let $G=(V, E)$ be a (connected)
graph with vertex set $V$ associated to the spins, and edge set $E$. Further,  let
$G^{\sigma}$ be an oriented version of $G$, where, for
every edge $e$, one end-vertex $v_e^+$ is assigned to be the
\emph{head} of $e$ , and the other end-vertex $v_e^-$ is the
\emph{tail} of $e$.
The Hamiltonian of the system is given by the general expression
\be
H(\{s_i\}):=\sum _{e \in E} h_e(|s_{v_e^+} -
s_{v_e^-}|_q).
\ee
Here every $h_e$ is a local Hamiltonian
defined on the edge $e$ with the only restriction that the
interaction strength between the spins on the endpoints
$v_e^+$ and $v_e^-$ is a function of the difference
between $s_{v_e^+}$ and $s_{v_e^-}$ modulo $q$, denoted by
$|s_{v_e^+} - s_{v_e^-}|_q$. Thus, a general inhomogeneous
model on an arbitrary graph is obtained.

Note that this general model specializes to known and widely studied
statistical mechanics models such as the standard Potts
model, the Ising model and the clock model \cite{Wu83}. For example, the $q$-state Potts model
(which specializes to the Ising model when $q=2$) is
obtained by taking $h_e(j):=-J_e\ \delta(j, 0)$, for every
$j\in\{0, \dots, q-1\}$ and for every $e\in E$, where the
$J_e$ are real coefficients.

The central quantity in this context is the partition function,
\be\label{partition}
Z_{G}(q,\sigma, \{h_e\}) := \sum_{\{s_i\}} e^{-\beta H(\{s_i\})},
\ee
where $\beta = (k_BT)^{-1}$, with $k_B$ the Boltzmann
constant and $T$ the temperature, from which other relevant physical quantities such as e.g. free energy can be derived.

\

%---------------------------------------------------
%---------------------------------------------------
{\bf 3.--- Quantum formulation.}
%---------------------------------------------------
%---------------------------------------------------
We now reformulate the partition function $Z_G$ in terms of
a quantum physics language. To this aim, we will associate
with the oriented graph $G^{\sigma}$ a quantum state of $N$
$q$--dimensional systems, where $N:=|E|$ is the total
number of edges of $G$, in the following way. Consider the
incidence matrix $B^{\sigma}$ of the oriented graph
$G^{\sigma}$ \cite{Go01}. This is the matrix with entries $0$ and $\pm 1$
where the rows are indexed by the vertices of $G$, and the
columns by the edges, such that $(B^{\sigma})_{a, e}=1$ if
the vertex $a$ is the head of the edge $e$,
$(B^{\sigma})_{a, e}=-1$ if $a$ is the tail of $e$, and
$(B^{\sigma})_{a, e}=0$ otherwise. Let $C_G(q, \sigma)$ be
the row space of $B^{\sigma}$ when arithmetic is performed
over $\mathbb{Z}_q$. This space is a linear subspace of
$\mathbb{Z}_q^{E}:=\bigoplus_{e\in E} \mathbb{Z}_q$. We now
associate to $C_G(q, \sigma)$ the quantum state \be
\label{CSS} |\psi_G\rangle:=\sum_{v\in
C_G(q,\sigma)}|v\rangle, \ee where the $|v\rangle$ are
basis states of $(\C^{q})^{\otimes N}$ obtained by taking
tensor products of a local basis $\{|0\rangle,\dots,
|q-1\rangle\}$. Note that we have omitted the dependence of
$|\psi_G\rangle$ on $q$ and $\sigma$ to keep notation
simple.

An interesting property is that the state $|\psi_G \rangle$ is a {\em stabilizer state}, i.e., it is the
unique joint eigenstate of $q^N$ commuting Pauli operators
\cite{Ho05, Sch00}. More specifically, defining the operators $X$ and $Z$
by
\be\label{XZ}
X|j\rangle = |j+1\ \mbox{mod }q\rangle,\quad
\mbox{ and }\quad Z|j\rangle = e^{\frac{2\pi ij}{q}}|j\rangle,
\ee
for every $j=0, \dots, q-1$, one finds that $|\psi_G\rangle$ is
invariant under the $q^N$ local operations \be X(u)Z(v):=\bigotimes_{e\in E} X^{u_e}Z^{v_e},\ee
for every $u\in C_G(q, \sigma)$ and $v\in C_G(q,
\sigma)^{\perp}$.
It is important to remark that the above construction of
associating a quantum stabilizer state $|\psi_G \rangle$ to
a graph $G$ is entirely different from the standard way of
defining a \emph{graph state}, usually denoted by
$|G\rangle$ \cite{He06}. While for $|\psi_G\rangle$ the
edges of $G$ are identified with quantum systems and the
incidence matrix of $G$ is used, yielding an $N$-particle
state (where $N = |E|$), in the construction of $|G\rangle$ the vertices are
identified with quantum systems and the adjacency matrix is
used, leading to a state on $n=|V|$ particles.

We can now relate the partition function $Z_G$ to the state
$|\psi_G\rangle$, and obtain the first main result of this
letter: \emph{The partition function $Z_{G}(q, \sigma,
\{h_e\})$ on an arbitrary graph $G$ with $N$ edges can be
written as the overlap between a quantum stabilizer state
on $N$ $q$-dimensional systems, and a complete product
state.} More precisely, we find \be \label{overlap}
Z_{G}(q,\sigma, \{h_e\}) = q\cdot \langle \psi_G|\left(
\bigotimes_{e\in E}|\alpha_e\rangle\right), \ee where
$|\alpha_e\rangle := \sum_{j=0}^{q-1} e^{-\beta h_e(j)
}|j\rangle$ is a $q$-dimensional vector associated to the
edge $e$, specifying the interaction \cite{Proof_overlap}.

Note that, whereas the states $|\psi_G\rangle$ encode interaction patterns, the product states $|\alpha_e\rangle$ encode the details of the interactions. For example, in the case of the Potts model one has \be |\alpha_e\rangle = e^{-\beta J_e}|0\rangle + (|1\rangle + \dots + |q-1\rangle).\ee

Examples of states $|\psi_G\rangle$ are obtained as follows.
First, if $G$ is a tree graph---for example, a 1D chain
with open boundary conditions---it is easy to show that
$C_{G}(q, \sigma)$ is the complete space $\mathbb{Z}_q^{E}$
(independent of the orientation $\sigma$). One then finds
that the corresponding quantum state is a product state,
namely $|\psi_{G}\rangle= |+\rangle^{\otimes N}$, where
$|+\rangle := |0\rangle +\dots + |q-1\rangle$. Hence, the
partition function $Z_G$ on trees can easily be evaluated,
and one has $Z_G\sim\prod_e\ \langle+| \alpha_e\rangle$.
Second, a 1D chain with periodic boundary conditions
corresponds for $q=2$ to a Greenberger-Horne-Zeilinger
(GHZ) state \be |\psi_{G}\rangle = |+\rangle^{\otimes N} +
|-\rangle^{\otimes N},\ee where $|-\rangle = |0\rangle
-|1\rangle$, and also this result leads to the well known
closed formula for the partition function. Finally, the 2D
rectangular lattice with open (periodic) boundary
conditions leads for $q=2$ to a \emph{planar (toric) code
state} as introduced in the context of topological quantum
error--correction \cite{Ki97}. We remark that the connection
between the 2D Ising model and
planar (toric) code states was first proven and utilized in Ref. \cite{Br06}.

\

%---------------------------------------------------
%---------------------------------------------------
{\bf 3.--- Solutions using quantum techniques.}
%---------------------------------------------------
%---------------------------------------------------
Having established the above formulation of the partition
function $Z_G$ in a quantum language, we now show how to
use powerful techniques established in quantum information
theory to study $Z_G$. Our general approach will be to
translate insights regarding the states $|\psi_G\rangle$
into insights regarding the partition functions $Z_G$. We
remark that it is not obvious that the formulation in a
quantum language simplifies the problem,  in particular as
it involves quantum states of $N$ $q$--dimensional systems,
which are described by $q^N$-dimensional vectors and thus
have an increased number of parameters. However, we can
make use of the following two advantageous features of the
correspondence (\ref{overlap}). First, as pointed out
above, the states $|\psi_G\rangle$ belong to the class of
stabilizer states, which are in general a highly manageable
family of multi-particle quantum states which has been
extensively studied, and of which many properties are
known. Second, it follows from (\ref{overlap}) that the
state $|\psi_G\rangle$ is only used to obtain an encoding
of the interaction pattern of the model---i.e., the
(oriented) graph $G$---and that all information regarding
the specifics of the interactions---i.e., the Hamiltonians
$h_e$---is encoded into the single particle states
$|\alpha_e\rangle$. Translating properties of the states
$|\psi_G\rangle$ into properties of the corresponding
partition functions thus yields a way of unifying
properties of models which are specializations of the
current model (such as the Ising, Potts and clock
models---see above), as one simply needs to consider
different local states $|\alpha_e\rangle$ for the same
state $|\psi_G\rangle$.

Next we give three illustrations how to use stabilizer
techniques to investigate the partition
function $Z_G$.

\

{\it 3.1. Efficient evaluation of $Z_G$.} First we consider
the problem of evaluating $Z_G$ in a given set of
hamiltonians $\{h_e\}$. This problem is known to be
NP--hard on general graphs. Efficient algorithms for
special instances of graphs do exist, e.g., for the Ising
model on planar graphs \cite{Ba82}. Here we show that:
\emph{The partition function $Z_G$ can be evaluated
efficiently (i.e., in polynomial time in the number of
edges $N$) on all graphs $G$ of logarithmically bounded
tree--width \cite{twd}. Moreover, we provide an  explicit
algorithm based on the efficient classical description of
quantum states in terms of tree tensor networks}.

To obtain this result, we use that any quantum state can be
represented by a tree tensor network (TTN) \cite{Shi05},
i.e., as a collection of tensors of rank $\chi_i$ that are
arranged according to a tree structure. In general, the
maximal rank $\chi:=\max \chi_i$ of the tensors in the
network grows exponentially with the number of systems $N$.
As shown in Ref. \cite{Shi05}, using the TTN description
one can extract information from the quantum state  with
overhead $O(N\cdot{\rm poly}(\chi))$. In particular, one
can calculate overlaps with product states. For stabilizer
states on qubits ($q=2$), it was shown how to obtain the
{\em optimal} TTN (i.e. with minimal $\chi$) in Ref.
\cite{Va06}, and this method can be extended to arbitrary
$q$ \cite{Va_in_prep}. It was shown in Ref. \cite{Va06}
that, for a given stabilizer state $|\psi\rangle$, the
optimal $\chi$ is obtained by computing the (exponential of
the) Schmidt--rank width of $|\psi\rangle$. Moreover, the
optimal TTN description of the state $|\psi\rangle$ can
explicitly be computed in  ${\rm poly (N,\chi)}$ steps on a
classical computer. Evaluation of overlaps with product
states, and, hence, specializing to the case of
$|\psi_G\rangle$, evaluation of partition functions $Z_G$
corresponds to a simple contraction of tensors, which can
be done in $O(N\cdot \chi^3)$ steps \cite{Shi05}. For
$|\psi_G\rangle$, one shows that the optimal $\chi$ scales
as the exponential of the tree--width $t(G)$ of the graph
$G$, by using that the Schmidt-rank width of
$|\psi_G\rangle$ is proportional to the branch-width $b(G)$
\cite{Hi03} of the underlying code $C_G(\sigma, q)$
\cite{Va_in_prep}; further, one uses the inequalities \be f_1(
b(G)) \leq t(G)\leq f_2 (b(G)),\ee where $f_1,f_2$ are linear functions \cite{Hi03}. It follows that whenever $t(G)$
grows at most logarithmically with $N$, then $\chi$ scales
polynomially with $N$, and hence the partition function
$Z_G$ can be evaluated {\em efficiently}.

Note that the computationally hardest part in the above algorithm is to obtain the TTN description of $|\psi_G\rangle$, while the calculation of overlaps with arbitrary product states, and hence the variation of coupling strengthes or different models for a fixed geometry, only scales linearly with $N$. We also emphasize that nonplanar graphs (of logarithmically bounded tree--width), as well as non--local interactions, are covered by this result. Results regarding efficient computation of homogeneous Potts model partition functions on graphs of (logarithmically) bounded tree--width have been obtained before \cite{An98}, though with entirely different methods. We emphasize that our approach, in contrast to previous approaches, can handle without difficulty also inhomogeneous models.
Moreover, our method leaves a lot of space for
generalizations.

\

{\it 3.2. Dualities for planar graphs.} Next we show how
the partition function $Z_G$ of a planar graph $G$ can be
related to the partition function $Z_D$ of its dual graph
$D$ \cite{Go01}. First, note that every orientation
$\sigma$ of $G$ induces an orientation of its dual $D$,
which we also denote by $\sigma$. We refer to Ref.
\cite{Go01} (p. 168) for details. Letting $B(G^{\sigma})$
and $B(D^{\sigma})$ be the incidence matrices of
$G^{\sigma}$ and $D^{\sigma}$, respectively, one then has
$B(G^{\sigma}) B(D^{\sigma})^T=0$ \cite{Go01} (p. 169),
which implies that the spaces $C_{G}(\sigma, q)$ and
$C_D(\sigma, q)$ are each other's duals, i.e.,
$C_{G}(\sigma, q)^{\perp}=C_D(\sigma, q)$. Now, let $F$ be
the quantum Fourier transform, or generalized Hadamard
operation, \be F:=\frac{1}{\sqrt{q}}\sum_{j, k=0}^{q-1}
e^{\frac{2\pi i}{q} kj} |j\rangle \langle k|,\ee defined on
$\mathbb{C}^q$. This operator satisfies $FXF^{\dagger} = Z$
and $FZF^{\dagger} = X$, where $Z$ and $X$ are defined in
(\ref{XZ}). Using this property and the duality of
$C_{G}(\sigma, q)$ and $C_D(\sigma, q)$, one finds that
%\be
\be |\psi_D\rangle = \left(\bigotimes_{e\in E}F \right)|\psi_G\rangle.\ee
%\ee
In other words, \emph{the states $|\psi_G\rangle$ and
$|\psi_D\rangle$ are equal up to a Hadamard operation
applied simultaneously on all local Hilbert spaces in the
system.} This result immediately relates the partition
functions $Z_G$ and $Z_D$, as one finds that \be \langle
\psi_G|\left(\bigotimes_{e\in E}|\alpha_e\rangle\right) =
\langle \psi_D|\left( \bigotimes_{e\in
E}|\alpha_e'\rangle\right),\ee where now $|\alpha_e'\rangle:=
H^{\dagger}|\alpha_e\rangle$. This leads to $Z_{G}(q,
\sigma, \{h_e\}) = Z_{D}(q, \sigma, \{h_e'\})$, where the
$h_e'$ are defined through \be\label{dual} e^{-\beta
h_e'(j)}:=\frac{1}{\sqrt{q}}\sum_{k=0}^{q-1} e^{\frac{-2\pi
i}{q} kj} e^{-\beta h_e(k)}, \ee for every $j=0, \dots,
q-1$.

As an example, consider the Potts model on the graph $G$,
which is obtained by putting $h_e(0):=-J_e$ and $h_e(j)=0$
whenever $j\neq 0$, for every $e\in E$. The identity
(\ref{dual}) then yields
\be
q^{1/2}e^{-\beta h_e'(j)} =
\left\{\begin{array}{ll} e^{\beta J_e} + q-1& \mbox{if
}j=0\\ e^{\beta J_e} -1&\mbox{if } j=1, \dots,
q-1.\end{array} \right.
\ee
Defining interaction strengths $J_e'$ through \be e^{\beta J_e'}:= \frac{e^{\beta J_e} +
q-1}{e^{\beta J_e} -1},\ee or, equivalently, $(e^{\beta J_e'}-1)(e^{\beta J_e}-1) = q$, one recovers the well know
high--low temperature duality relation for the Potts model
partition function \cite{Wu83}, where the partition function on the
graph $G$ with interaction strengths $\{J_e\}$ is related
to the partition function on the dual graph $D$ with
interactions $\{J_e'\}$.

\

{\it 3.3. Symmetries.} Finally, general symmetries can
easily be obtained. Here the essential observation is the
following: any local unitary operator $U:= \bigotimes_{e\in
E} U_e$ having $|\psi_G\rangle$ as an eigenstate with
eigenvalue $\lambda$, yields a symmetry \be \langle
\psi_G|\left(\bigotimes_{e\in E}|\alpha_e\rangle\right) =
\langle \psi_G|\left( \bigotimes_{e\in
E}|\alpha_e'\rangle\right), \ee where $|\alpha_e'\rangle:=
\bar\lambda U_e|\alpha_e\rangle$. A symmetry is obtained in
the sense that different configurations of interactions
(which are encoded in the $|\alpha_e\rangle$) on the same
graph $G$ yield the same partition function\cite{complex}.
Here again the fact that the states $|\psi_G\rangle$ are
stabilizer states is particularly advantageous, as every
$|\psi_G\rangle$ is the joint eigenstate of the $q^{N}$
local Pauli operators $X(u)Z(v)$ (see (\ref{XZ}) and
below), which constitute the stabilizer group of
$|\psi_G\rangle$. Thus, this stabilizer group corresponds
to a group of $q^N$ symmetries of $Z_G$.

Let us give an example of such a symmetry for $q=2$. Let
$G$ be an arbitrary graph, let $a$ be one of its vertices,
and let $E(a)$ be the set of edges incident with $a$. Let
$X[a]$ be the $N$-qubit correlation operator on such  that
the the $e$th tensor factor is equal to the Pauli matrix
$X$ if $e\in E(a)$, and equal to $I$ otherwise. Then $X[a]$
stabilizes $|\psi_{G, 2}\rangle$ and, using
the relation with $Z_G$, one obtains a relation
between $Z_G(2; \{J_e\})$ and $Z_G(2; \{J_e'\})$, where
$J_e':=-J_e$ if $e\in E(a)$, and $J_e':=J_e$ otherwise.

\

%---------------------------------------------------
%---------------------------------------------------
{\bf 4.--- Relation to measurement based quantum computation.}
%---------------------------------------------------
%---------------------------------------------------
So far we have used stabilizer  techniques to study the
partition function $Z_G$, using the relation
(\ref{overlap}). Here we remark that this connection can
also naturally be invoked to gain insights in the opposite
direction---i.e., insights in $Z_G$ can be used to
understand the properties of the states $|\psi_G\rangle$
and their possible role in quantum information tasks.  In
particular, one may consider the question whether the
states $|\psi_G\rangle$ are useful resources for
measurement based quantum computation (MQC), similar to the
2D cluster state, which in fact allows universal
computation by local measurements only \cite{Ra01}. Like
all stabilizer states, the $|\psi_G\rangle$ are natural
candidates for resources for MQC, as these states are
generally highly entangled and as they can be produced
efficiently by a poly-sized quantum circuit. The connection
with the partition function immediately {\em relates the
difficulty of evaluating $Z_G$ with the difficulty of
classically simulating local measurements on the state
$|\psi_G\rangle$}. In particular, one finds that all
(classes of) graphs for which the computation of $Z_G$ is
hard, give rise to resource states for which simulation of
MQC is a hard problem. For example, nonplanar graphs of
large tree-width would typically give rise to states where
MQC is NP-hard to simulate classically. Conversely, known results regarding efficient evaluation of $Z_G$ on certain graphs---e.g., planar graphs for the Ising model ($q=2$)---may lead to states $|\psi_G\rangle$ on which MQC can be simulated efficiently classically. We refer to Ref. \cite{Br06}, where this approach was first adopted to show that MQC on Kitaev's planar code states can be simulated efficiently classically.

\

%---------------------------------------------------
%---------------------------------------------------
{\bf 5.--- Summary and outlook.}
%---------------------------------------------------
%---------------------------------------------------
We have related the  evaluation of the partition function
on a general class of classical spin models to the
calculation of overlaps between quantum stabilizer states
encoding the interaction pattern, and product states
encoding the details of the interaction. This allowed us to
use powerful techniques established in quantum information
theory to obtain novel classical algorithms to evaluate the
partition function, and to investigate general symmetries
and duality relations. We emphasize that in all cases we
consider, the reformulation in terms of a quantum problem
is virtual, in the sense that we provide an alternative way
to obtain a classical solution of the initial classical
problem by making use of techniques established to describe
and simulate quantum systems by classical means.

Finally, we note that other  classical combinatorial
problems can be formulated in the language of quantum
physics using a similar construction as outlined above. As
an example, we mention that the weight enumerator of a
classical code---a central quantity in classical coding
theory---can be evaluated with similar methods
\cite{Va_in_prep}. Further, as the Potts model partition
function is intimately related to problems in graph and
knot theory through the relation with the Tutte polynomial,
the present results also have implications in these fields.

\begin{acknowledgements}
We thank R. Raussendorf, S. Bravyi and H. Wagner for interesting discussions. Work
supported by the FWF, the European Union (QICS, OLAQUI,
SCALA), and the \"OAW through project APART (W.D.).
\end{acknowledgements}


\begin{thebibliography}{99}
\bibitem{twd}
The tree-width of a graph is a parameter which measures how similar a graph is to a tree \cite{Ro84}.
E.g., the tree width of a tree is equal to one; the tree width of a $k\times k$ square lattice is $O(k)$. Examples of graphs with logarithmically bounded tree width include all $k\times l$ strips of 2D (planar or nonplanar) lattices, where $k=O(\log l)$.
\bibitem{Proof_overlap}
To prove (\ref{overlap}), first note
that $ \langle i | \alpha_e\rangle = \exp({-\beta h_e(i)
}).$ Further, every $y\in C_G(q, \sigma)$ can be written as
$ y=(B^{\sigma})^Ts$, where $s\in \mathbb{Z}_q^{n}$, where
$n$ is the number of vertices of $G$, and therefore $y_e =
|s_{v_e^+} - s_{v_e^-}|_q$ for every $e$. The factor $q$
in (\ref{overlap}) is due to the fact that the rows of
$B^{\sigma}$ are not linearly independent, as the sum of
all rows yields the zero vector. Therefore, there are
exactly $q$ ways of writing $y$ as a linear combination of
the rows of $B^{\sigma}$, and one has to divide by $q$ to
account for this overcounting.
\bibitem{complex}
Note that $U_e$ may take
$|\alpha_e\rangle$, which has all real and positive coefficients,
to a complex vector.
\end{thebibliography}
\end{document}